# Nonlinear mixed-effect models for prostate-specific antigen kinetics and link with survival in the context of metastatic prostate cancer: a comparison by simulation of two-stage and joint approaches

Solène Desmée[a,b,*], France Mentré[a,b], Christine Veyrat-Follet[c], Jérémie Guedj[a,b]

[a] INSERM, IAME, UMR 1137, F-75018 Paris, France.

[b] Université Paris Diderot, IAME, UMR 1137, Sorbonne Paris Cité, F-75018 Paris, France.

[c] Drug Disposition, Disposition Safety and Animal Research Department, Sanofi, Chilly-Mazarin, France

[*] To whom correspondence should be addressed (e-mail: Solene.desmee@inserm.fr)

Running head : NLMEM for PSA kinetics and link with survival

*Abstract*

In metastatic castration-resistant prostate cancer (mCRPC) clinical trials, the assessment of treatment efficacy essentially relies on the time-to-death and the kinetics of prostate-specific antigen (PSA). Joint modelling has been increasingly used to characterize the relationship between a time-to-event and a biomarker kinetics but numerical difficulties often limit this approach to linear models. Here we evaluated by simulation the capability of a new feature of the Stochastic Approximation Expectation-Maximization algorithm in Monolix to estimate the parameters of a joint model where PSA kinetics was defined by a mechanistic nonlinear mixed-effect model. The design of the study and the parameter values were inspired from one arm of a clinical trial. Increasingly high levels of association between PSA and survival were considered and results were compared with those found using two simplified alternatives to joint model, a two-stage and a joint sequential model. We found that joint model allowed for a precise estimation of all longitudinal and survival parameters. In particular the effect of PSA kinetics on survival could be precisely estimated, regardless of the strength of the association. In contrast, both simplified approaches led to bias on longitudinal parameters and two-stage model systematically underestimated the effect of PSA kinetics on survival.

In summary we showed that joint model can be used to characterize the relationship between a nonlinear kinetics and survival. This opens the way for the use of more complex and physiological models to improve treatment evaluation and prediction in oncology.

# Introduction

Prostate cancer is the second most frequently diagnosed cancer in men and is responsible for about 300 000 deaths worldwide every year [1]. Although treatment can be effective, a number of factors, such as resistance or delayed treatment (4% of cancer have metastasized at diagnosis [2]) considerably worsen the treatment outcome. In the case of metastatic castration-resistant prostate cancer (mCRPC), the evaluation of chemotherapy efficacy primarily relies on the overall survival [3] and is complemented by the analysis of the Prostate-Specific Antigen (PSA). Although countless studies have explored the relationship between different PSA kinetic parameters (such as doubling time or time to nadir) and survival, the choice of a clearly-defined parameter remains controversial.

The lack of consensus on how to use PSA kinetics is exacerbated by the difficulty to properly handle PSA kinetics and the time-to-event (time-to-death or dropout) in statistical models. Several methods have been proposed in the literature. The simplest one is to plug the individual PSA kinetic parameters into a survival model [4]. However the fact that these parameters are often not directly observed from the data makes this approach error prone. A second approach is to use a model to describe the entire PSA kinetics using mixed-effect models to precisely account for between-subjects variability, and then to plug individual model predictions as covariates in a survival model [5,6]. However this method, called in the following "two-stage" approach, is prone to bias because it does not take into account the relationship between the marker's kinetic and the time-to-event and the uncertainty in the individual model predictions [7]. In order to eliminate the bias found in the two-stage approach, one can use models which simultaneously, or "jointly", handle longitudinal and time-to-event data [7–13]. For the latter, one can either aim to estimate all parameters simultaneously ("joint" model) or in a sequential manner ("joint sequential" model), as it has been suggested in the pharmacometric field [14].

The main challenge in using joint model is the numerical complexity involved by the calculation and the maximization of the likelihood. This difficulty can be circumvented by using a linear model for the PSA kinetics as implemented in the large majority of published models [7,15] or softwares [16]. However this precludes the use of physiologically more accurate models for PSA kinetics which are in essence nonlinear.

In the last years, pharmacometric softwares have addressed the need for joint model when the longitudinal model is defined by a nonlinear mixed-effect model (NLMEM). This approach was firstly implemented in NONMEM using a Laplacian approximation of the likelihood and was essentially used to account for informative dropouts [17,18]. Recently the Stochastic Approximation Expectation-Maximization (SAEM) algorithm, a method relying on an "exact" calculation of the likelihood, was extended to include time-to-event data [19,20] in Monolix and NONMEM.

Here we evaluated the benefit of joint models using the SAEM algorithm for characterizing the relationship between survival and a biomarker having a non-linear kinetics. We compared by simulation the precision and the type 1 error of longitudinal and survival parameters obtained using a joint, a joint sequential and a two-stage model in the context of a clinical study in mCRPC according to the strength of the association between PSA kinetics and survival.

# Models and notations

## *A Mechanistic model for PSA kinetics*

In absence of treatment we assume that prostatic cells, C (mL$^{-1}$), proliferate with rate r (day$^{-1}$) and are eliminated with rate d (day$^{-1}$). PSA (ng.mL$^{-1}$) is secreted with a production rate p (ng.day$^{-1}$) and cleared from the blood with rate δ (day$^{-1}$). We suppose that a chemotherapy for mCRPC acts by blocking cell proliferation with time-varying effectiveness, e(t), and hence the proliferation rate under treatment is given by $r' = r(1 - e(t))$ with $0 \leq e(t) \leq 1$ (Figure 1):

$$\begin{cases} \frac{dC}{dt} = r(1 - e(t))C(t) - dC(t) \\ \frac{dPSA}{dt} = pC(t) - \delta PSA(t) \end{cases} \quad (1)$$

Treatment is initiated at t=0, PSA(0)=PSA$_0$ and C(0)=C$_0$ are PSA value and the number of prostatic cells at treatment initiation respectively. Here, treatment is assumed to be constantly effective until a certain time, T$_{esc}$, after what tumor escapes and treatment has no longer an effect:

$$e(t) = \begin{cases} \varepsilon & if\ t \leq T_{esc} \\ 0 & if\ t > T_{esc} \end{cases} \quad (2)$$

Further, we made the assumption of quasi steady-state at treatment initiation and thus $C_0 = \frac{\delta \times PSA_0}{p}$. With this piecewise constant treatment effect (2), the model (1) has an analytical solution given by:

$$PSA(t, \psi) =$$
$$\begin{cases} \frac{\delta PSA_0}{r(1-\varepsilon)-d+\delta} e^{(r(1-\varepsilon)-d)t} + \left[PSA_0 - \frac{\delta PSA_0}{r(1-\varepsilon)-d+\delta}\right] e^{-\delta t} & if\ t \leq T_{esc} \\ \frac{\delta PSA_0}{r-d+\delta} e^{(r-d)t - r\varepsilon T_{esc}} + \left[PSA(T_{esc}) - \frac{\delta PSA_0 e^{(r(1-\varepsilon)-d)T_{esc}}}{r-d+\delta}\right] e^{-\delta(t-T_{esc})} & if\ t > T_{esc} \end{cases} \quad (3)$$

Because only 4 parameters can be identified from Eq. 3, we fixed d to 0.046 day$^{-1}$, corresponding to a half-life of tumor cells of 15 days, consistent with an apoptotic index of 5% in metastatic prostate cancer [21]. Moreover we fixed δ to 0.23 day$^{-1}$, corresponding to a PSA half-life in blood of about 3 days [22]. Thus PSA kinetics was defined by the vector parameter ψ={r, PSA$_0$, ε and T$_{esc}$}.

## *Statistical model for PSA measurements*

Let *y$_{ij}$* denote the *j$^{th}$* longitudinal measurement of log(PSA+1) for the individual *i* at time *t$_{ij}$*, where *i=1, …, N, j=1, …, n$_i$*, N is the number of subjects and *n$_i$* the number of measurements in subject *i*. The nonlinear mixed-effect model (NLMEM) for PSA is defined as follows:

$$y_{ij} = \log(PSA(t_{ij}, \psi_i) + 1) + \sigma e_{ij} \quad (4),$$

where PSA is given by the formula (3), ψ$_i$ is the vector of the individual parameters and e$_{ij}$ the residual error which follows a standard normal distribution with mean 0 and variance 1. The vector of the individual parameters ψ$_i$ is decomposed as fixed effects μ={r, PSA$_0$, ε, T$_{esc}$} representing median effects of the population and random effects η$_i$ specific for each individual. It is assumed that η$_i$~N(0, Ω) with Ω the variance-covariance matrix. In this work, Ω is supposed to be diagonal. Each diagonal element $\omega_q^2$ represents the variance of the $q^{th}$ component of the random effect vector $\eta_i$. We assumed log-normal distribution for r, $PSA_0$ and $T_{esc}$:

$$\log(\psi_{q,i}) = \log(\mu_q) + \eta_{q,i}$$

and logit-normal distribution for ε:

$logit(\varepsilon_i) = logit(\mu_\varepsilon) + \eta_{\varepsilon,i}$ with $logit(x) = \log\left(\frac{x}{1-x}\right)$ for 0<x<1.

The population parameters vector of PSA noted $\theta_l$ is composed of {μ, Ω, σ}.

### *Statistical model for survival*

Let $X_i$ denote the time to event and $C_i$ the censoring time for the $i^{th}$ subject. Survival data ($T_i$, $\delta_i$) are observed in all individuals, where $T_i$=min($X_i$, $C_i$) and $\delta_i = 1, \text{if } X_i \leq C_i$, 0 otherwise. For the event process, we used a hazard function of the form:

$$h_i(t|PSA(t,\psi_i)) = h_0(t)\exp(\beta PSA(t,\psi_i)) \quad (5)$$

where the baseline hazard function, $h_0$, is a Weibull hazard function $h_0(t) = \frac{k}{\lambda}\left(\frac{t}{\lambda}\right)^{k-1}$. The parameter β measures the strength of the association between the PSA kinetics and the risk of death. If β=0, the survival process is independent on the PSA evolution and survival data can be fitted by a Weibull model without adjusting for PSA. If β≠0, the survival process depends on the PSA kinetics. The survival parameters to estimate are $\theta_s$={λ, k, β}.

### *Joint model*

Joint models assume conditional independence: given the random effects $\eta_i$, longitudinal measurements and survival events are independent. All parameters, $\theta = \{\theta_l, \theta_s\}$ are simultaneously estimated. Thus the joint log-likelihood for subject *i* can be written as follow [23]:

$$l_i(\theta) = \log \int p(y_i|\eta_i;\theta)\{h_i(T_i|\eta_i;\theta)^{\delta_i}S_i(T_i|\eta_i;\theta)\}p(\eta_i;\theta)d\eta_i \quad (6)$$

where $S_i(t|\eta_i;\theta) = \exp\left(-\int_0^t h_0(s|\eta_i;\theta)\exp(\beta PSA(s,\psi_i))ds\right)$ is the survival function conditionally on the random effects, $p(y_i|\eta_i;\theta)$ the density of the longitudinal observations conditionally on the random effects and $p(\eta_i;\theta)$ the density of the random effects.

## *Two-stage model*

In order to simplify the heavy calculations involved by Equation (6) one may also use a two-stage approach. In the first step PSA kinetics parameters (θ$_l$) are estimated assuming complete independence of PSA kinetic and survival and the Empirical Bayes Estimates (EBEs), defined as the mode of the conditional distribution $p(\psi_i|y_i,\hat{\theta}_l)$, are used to provide individual parameters, noted $\hat{\psi}_i$. In the second step, the survival parameters (θ$_s$) are estimated maximizing the usual log-likelihood $l(\theta_s) = \sum_{i=1}^n \log\left\{h_i(T_i|PSA(t,\hat{\psi}_i);\theta_s)^{\delta_i} S_i(T_i|PSA(t,\hat{\psi}_i);\theta_s)\right\}$.

This method is analogous to the sequential method called "Individual Pharmacokinetic Parameters" (IPP) used in pharmacometric field to handle PharmacoKinetic/PharmacoDynamic (PKPD) data [14].

## *Joint sequential model*

In order to reduce the number of parameters to estimate in Equation (6), an alternative consists in first estimating population PSA kinetic parameters, as done in the two-stage model, and then estimating parameters related to survival (θ$_s$) fixing the all population PSA parameters (θ$_l$) in Equation (6) to the values obtained at the previous stage. This method, inspired from the "Population PK parameter and individual PK data" for the combined analysis of PKPD data, has been shown to limit the bias compared to two-stage approach [14].

# Simulation study

## *Design*

Simulation setting was inspired by one arm of a phase III study of clinical development of chemotherapy for metastatic prostate cancer [3]. M=100 datasets with N=500 patients were simulated with PSA measurements every 3 weeks for 2 years (i.e., the last possible measurement time was at t=735 days), leading to a maximum of 36 observations per patient (Table I). Follow-up was censored at t=735 and no other mechanism than death was considered for dropout. For the simulation of the time to death, k was fixed to 1.5, and 3 values for β were considered: 0, 0.005 and 0.02, corresponding to 'no link', 'low link' and 'high link' between PSA kinetics and survival, respectively. In order to maintain a comparable amount of PSA measurements across scenarios, λ was determined in each scenario such that the probability of survival at the end of the follow-up (i.e., 735 days) was equal to 25% with the median PSA kinetic parameters (Table I), leading to λ=580, 765 and 2150 in scenarios 'No link', 'Low link' and 'High link', respectively (Table II). Lastly, we evaluated the effects of having both a large baseline risk and a strong effect of PSA kinetics by setting β=0.02 and λ=580 in the scenario called 'Short survival'. Figure 2 shows the survival functions for these 4 scenarios for the "median patient", i.e., a virtual patient having PSA kinetic parameters equal to the median values in the population.

In order to take into account the effect of withdrawals from PSA follow-up, we also explored additional scenarios where PSA and/or vital status were censored in case of PSA progression defined as an increase of 25% above the nadir and of at least 2 ng/ml compared to nadir (see supplementary file 1).

All simulations were carried out with the R software version 3.0.1 [24].

## *Parameter estimation*

The estimation of two-stage, joint sequential and joint models was performed for each dataset and each scenario using the Stochastic Approximation Expectation-Maximization (SAEM) algorithm in the software Monolix version 4.2.2.

The initial values of the parameters were those used for the scenario 'No link'. Minus twice log-likelihood (-2LL) was calculated by Importance Sampling, fixing the Monte-Carlo sizes to achieve a sufficient level of precision for Likelihood Ratio Test (LRT) to 100 000 Monte-Carlo sizes for the scenario 'No link', 20 000 for the other scenarios in joint model and 2 000 in two-stage model.

SAEM was run using 1 chain and we verified that using 3 chains, as suggested in a related context [19], gave similar results. CPU (Central Processing Unit) times for parameters estimations and -2LL estimations were recorded on a i7 64bits 3.33GHz.

*Evaluation criteria*

We used Relative Estimation Errors defined by: $REE(\hat{\theta}_m) = \frac{\hat{\theta}_m - \theta^*}{\theta^*} \times 100$, where $\theta^*$ and $\hat{\theta}_m$ are the true and estimated parameters, respectively, for dataset m, m=1…M. Boxplots of the REEs with the 10% and 90% percentiles were plotted. When β=0 (scenario 'No link'), the boxplot of the absolute estimation error, $\hat{\beta}_k$, with the 10% and 90% percentiles were plotted .

The type 1 error and power were calculated as the proportion among the M datasets for which LRT (called in the following "uncorrected LRT") led to reject the null hypothesis $H_0$: β=0. The type 1 error was evaluated in the scenario 'No link' and the power was calculated for the 3 other scenarios. The significance level of the tests α for the observed type 1 error was fixed to 0.05, leading to a 95% prediction interval for 100 replicates equal to [0.7%-9.3%].

In some cases the estimation of -2LL, which relies on a stochastic approximation, was associated with a non-negligible standard error. Because this can lead to an inflation of the type 1 error, we also evaluated a "corrected LRT". In this test $H_0$ was rejected if $-2LL(H_0) + 2LL(H_1)$ was larger than $(\chi_1^2 + z_\alpha \sqrt{se_{-2LL}(H_0)^2 + se_{-2LL}(H_1)^2}\ ;\ \infty)$ where $se_{-2LL}(H_0)$ and $se_{-2LL}(H_1)$ are the estimated standard error of -2LL under $H_0$ and $H_1$, respectively, whereas $\chi_1^2$ and $z_\alpha$ are the chi-squared value with 1 degree of freedom and the $\alpha^{th}$ quantile of the standard normal distribution, respectively. $\sqrt{se_{-2LL}(H_0)^2 + se_{-2LL}(H_1)^2}$ is the uncertainty for the sum of the two estimated -2LL. We used a corrected LRT if $z_\alpha \sqrt{se_{-2LL}(H_0)^2 + se_{-2LL}(H_1)^2}$ was non negligible compared to $\chi_1^2$, i.e., for a significance level of 5%, $se_{-2LL}(H_0)^2$ or $se_{-2LL}(H_1)^2$ non negligible compared to 2.

## Results

### *Simulated data*

Figure 3 shows the spaghetti-plots along with the Kaplan-Meier curves of one typical dataset for each of the 4 scenarios. Because the scenario 'No link' assumes that death does not depend on PSA kinetics, PSA rebound after loss of treatment efficacy (at time $T_{esc}$) was more frequently observed than in the 3 other scenarios. As expected (see methods), the numbers of measurements in the first 3 scenarios were largely similar (Table III). In the last scenario where both the baseline hazard function and the effect of PSA kinetics were large, early events frequently occurred and the total amount of PSA data was much smaller.

### *Estimation performance*

No bias was observed in the scenario 'No link' regardless of the approach. In the two-stage approach, increasing effect of PSA on survival led to higher levels of bias (Figures 4 and 5). In particular there was a systematical underestimation of the PSA effect on survival with median (Q1;Q3) REE for β equal to -8.4% (-12.6 ; -4.1), -14.8% (-18.9 ; -10.2) and -9.1% (-16.7 ; -5.7) in scenarios 'Low link', 'High link' and 'Short survival', respectively. The bias was corrected by using joint sequential models or joint models, with median (Q1;Q3) REE for β of 0.005% (-3.6 ; 5.3), -0.6% (-3.3 ; 2.9) and 0.03% (-4.8 ; 5.7) in scenarios 'Low link', 'High link' and 'Short survival', respectively for this latter method. Although the three approaches led to low REEs for the almost all parameters of PSA kinetics (Figure 4), a bias was observed for the proliferation rate of tumor cells, r, with median (Q1;Q3) REE equal to -0.4% (-0.7 ; -0.05), -0.8% (-1.0 ; -0.5) and -1.2% (-1.8 ; -0.7) in scenarios 'Low link', 'High link' and 'Short survival', respectively, when using the two-stage or joint sequential approach. Here as well, the bias was largely reduced when using a joint model, with median (Q1;Q3) REE for r equal to 0.01% (-0.4 ; 0.3), -0.004% (-0.3 ; 0.2) and -0.08% (-0.6 ; 0.5), in scenarios 'Low link', 'High link' and 'Short survival', respectively. Of note large REE were found for the parameters λ and k (|REE|>30%) in the scenario 'High link' when using joint and joint sequential models, due to the presence of a local extremum of the likelihood function.

Lastly we also considered additional scenarios where PSA data and/or vital status were censored in case of PSA increase (see supplementary file 1). Although the precision of the parameter estimates was deterioriated due to the reduction in the amount of data available no substantial bias in the PSA kinetic parameters was found. However bias was found in survival related parameters, in particular when both PSA and vital status were censored after disease progression and this bias systematically led to an overestimation of the hazard function.

## Test performance

The uncorrected LRT led to a type 1 error of 21%, 9% and 12% for the joint model, the joint sequential model and the two-stage model, respectively, i.e., outside the 95% prediction interval ([0.7%-9.3%] for 100 replicates). The use of a corrected LRT (see methods) led to a smaller type 1 error of 4%, 3% and 12% for the joint model, the joint sequential model and the two-stage model, respectively. The reason why the type 1 error for the two stage model with or without correction were similar is due to the fact that the standard error of -2LL were negligible with this model ($<10^{-4}$). For the scenarios with β≠0, the power was 100% with all three models, regardless of the correction.

## Computation time

Mean CPU times for the simultaneous estimation of the 12 parameters using joint model were about 5 times larger than the total CPU times using two-stage model (70 vs 14 minutes) and about 1.2 times larger than the total CPU times using joint sequential model (70 vs 59 minutes), ignoring time for specific data manipulation required for setting the two-stage and joint sequential approaches. Mean CPU times for the -2LL estimation using joint model (respectively joint sequential model) and 100,000 chains was about 3.1 (respectively 3.6) times larger than when using the joint approach and 20,000 chains (264 vs 86 minutes (respectively 207 vs 57 minutes)) but led to a mean $se_{-2LL}(H_1)$ of 1.96 vs 5.14 (respectively 2.40 vs 5.90).

# Discussion

Numerical complexities have long limited the use of joint models to longitudinal processes defined by linear mixed-effect models [7,10,15]. Here we evaluated by simulation, in the context of PSA and survival in metastatic prostate cancer, the capability of a new feature of the Stochastic Approximation Expectation-Maximization algorithm in Monolix to estimate the parameters of a joint model where the longitudinal process was defined by a nonlinear model. And we compared the results to two simplified approaches, two-stage and joint sequential models.

We found that joint model provided unbiased parameters of both longitudinal and survival processes. In contrast, the use of a two-stage model [5,25] led to large biases when PSA kinetics and survival were linked. In particular, the impact of the biomarker kinetics on the survival, as measured by the link parameter $\beta$, was systematically underestimated, consistent with results found in linear mixed-effect model [8,9]. Beside survival parameters, a bias on the tumor proliferation rate, r, which is the driving force for the increase in PSA, was also observed in scenarios with $\beta \neq 0$. The fact that no such bias occurred when using joint model shows that the simultaneous estimation of the hazard function also improved the estimation of the longitudinal parameters. Moreover, a two-stage approach led to an inflation of type 1 error (i.e., conclude to an effect of PSA on survival while there is none) which could be explained by the shrinkage of the EBEs in patients with short survival [26]. By construction the joint sequential model led to the same biases on PSA longitudinal parameters than the two-stage model, but the estimation of survival parameters was close to that obtained with the joint model. This, therefore, suggests that joint sequential model could be a relevant approach when joint model cannot be performed.

In spite of the increasing numerical capability, the likelihood of joint models remains particularly complex to calculate. Here we reported that the likelihood was estimated with a relatively large uncertainty. Increasing the Monte-Carlo sizes led to smaller standard errors of the likelihood but considerably increased the computation time. The impact of this error on type 1 error was in part accounted by using a corrected likelihood ratio test (LRT). However more studies will be needed to precisely determine when this correction is necessary and whether it can be improved. Here for instance the corrected rejection region did not take into account the covariance between the likelihood calculated under the null and alternative hypotheses. Although the Wald test may be an alternative, we found here that standard errors of the estimates tended to be smaller than the Root Mean-Squared Errors (RMSE), indicating a potential underestimation of the standard errors. Furthermore, and in spite of the stochastic algorithm, the estimation of parameters related to survival was complicated in some cases by the existence of local extremum of the likelihood function. This stresses the need, in practice, to perform several runs of likelihood maximization using different initial values.

The main advantage of nonlinear models is the possibility to develop physiological models based on nonlinear differential equations, which naturally integrate the correlations between the different biomarkers. In this study, we used a rather simple model where the treatment effectiveness was piecewise constant, which allowed us to derive a biexponential analytical solution for the PSA kinetics. However this model may clearly be over-simplistic and more complex models will be needed that rely on several markers, such as tumor size or drug pharmacokinetics. The facilitated use of these models via joint models holds the promise that the determinants of survival may be much better characterized [20].

Regarding the survival model, we restricted here to a rather simple fully parametric model, where the baseline hazard function was a Weibull model [27] and the hazard function was related to the current PSA values. In practice complex survival models could be evaluated and standard tools for model selection (e.g., AIC or BIC) could be used to evaluate the effect of various transformations of the biomarker, such as the derivative or the cumulative value of PSA. Lastly with a fully parametric model the prediction and the simulation of the individual hazard function can easily be performed, making possible to guide treatment adaptation in a dynamic manner [28].

## Conclusion

SAEM algorithm implemented in Monolix was shown to provide precise estimates for joint models where the longitudinal model was defined by a non-linear mixed-effect model. This opens the way for a more systematic use of joint models and a better understanding of the relationship between biomarker kinetics and survival, especially in the field of metastatic cancer where survival and non-linear biomarker kinetics are intrinsically related.

*Acknowledgements*: The authors would like to thank Drug Disposition Department, Sanofi, Paris which supported Solène Desmée by a research grant during this work. We also thank Hervé Le Nagard for the use of the computer cluster services hosted on the "Centre de Biomodélisation UMR1137".

Table I: Values of the population PSA parameters used for the simulations in all scenarios

|  | Fixed effects | Transformation | Inter-individual standard deviation ($\omega$) |
|---|---|---|---|
| r (day$^{-1}$) | 0.05 | log-normal | 0.1 |
| $PSA_0$ (ng.mL$^{-1}$) | 80 | log-normal | 0.6 |
| $\varepsilon$ | 0.3 | logit-normal | 1.5 |
| $T_{esc}$ (day) | 140 | log-normal | 0.6 |
| $\sigma$ | 0.36 | - | - |

Table II: Values of the population survival parameters used for the simulations of the 4 scenarios

|  | Scenario No link | Scenario Low link | Scenario High link | Scenario Short survival |
|---|---|---|---|---|
| $\beta$ | 0 | 0.005 | 0.02 | 0.02 |
| $\lambda$ (day) | 580 | 765 | 2150 | 580 |
| k | 1.5 | 1.5 | 1.5 | 1.5 |

Table III: Number of PSA measurements per patient and median survival in the total number of simulated (50000) patients for the 4 simulated scenarios

| Number of PSA measurements | Scenario No link | Scenario Low link | Scenario High link | Scenario Short survival |
|---|---|---|---|---|
| 1-5 | 7% | 7% | 8% | 29% |
| 6-10 | 12% | 13% | 10% | 20% |
| 11-20 | 26% | 31% | 22% | 26% |
| 21-35 | 30% | 29% | 21% | 17% |
| 36 | 24% | 21% | 39% | 8% |
| Median survival (day) | 457 | 414 | 552 | 217 |

Figure 1: Schema of the secretion of PSA by prostate and cancer cells. PSA is expressed in ng.mL$^{-1}$ and C in mL$^{-1}$. r is the rate of prostatic cells proliferation in absence of treatment (day$^{-1}$), d the rate of prostatic cells elimination (day$^{-1}$), p the rate of PSA secretion by C (ng.day$^{-1}$), δ the rate of PSA elimination (day$^{-1}$) and e(t) the time-dependent treatment effect.

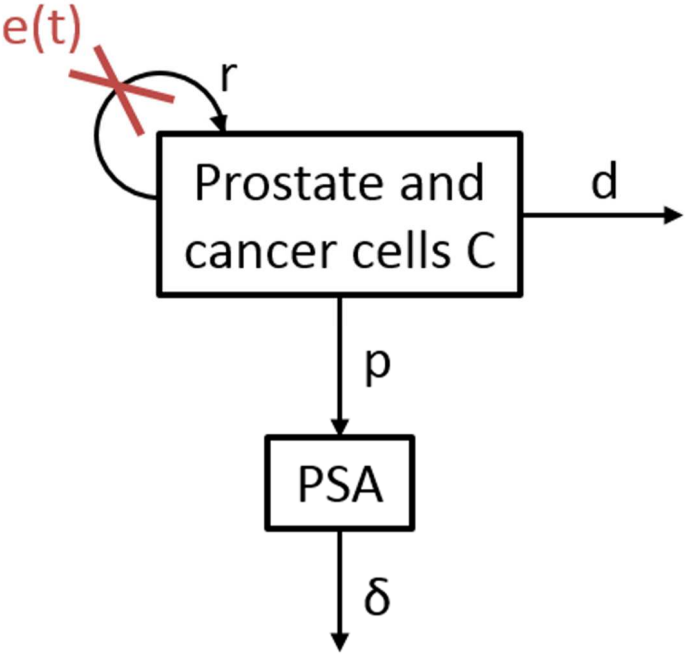

Figure 2: Typical evolution of PSA(t) (solid black) and survival functions in the typical patient (who have the fixed effects of the Table I as PSA parameter) for the scenarios 'No link' (β=0, λ=580) (solid orange), 'Low link' (β=0.005, λ=765) (dashed green), 'High link' (β=0.02, λ=2150) (dotted blue) and 'Short survival' (β=0.02, λ=580) (dotdash purple)

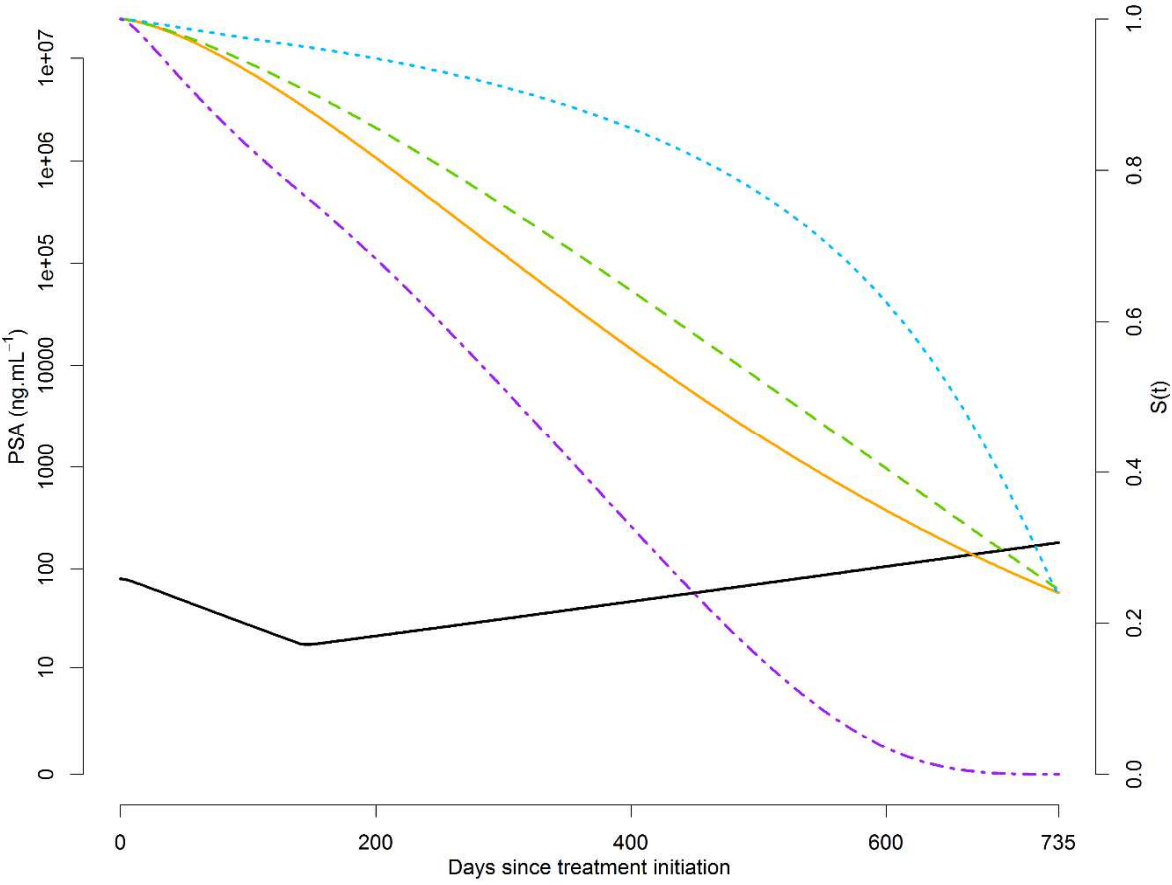

Figure 3: Spaghetti-plots (black lines) and estimated Kaplan-Meier curves (colored lines) for one typical dataset (N=500) for each of the four scenarios

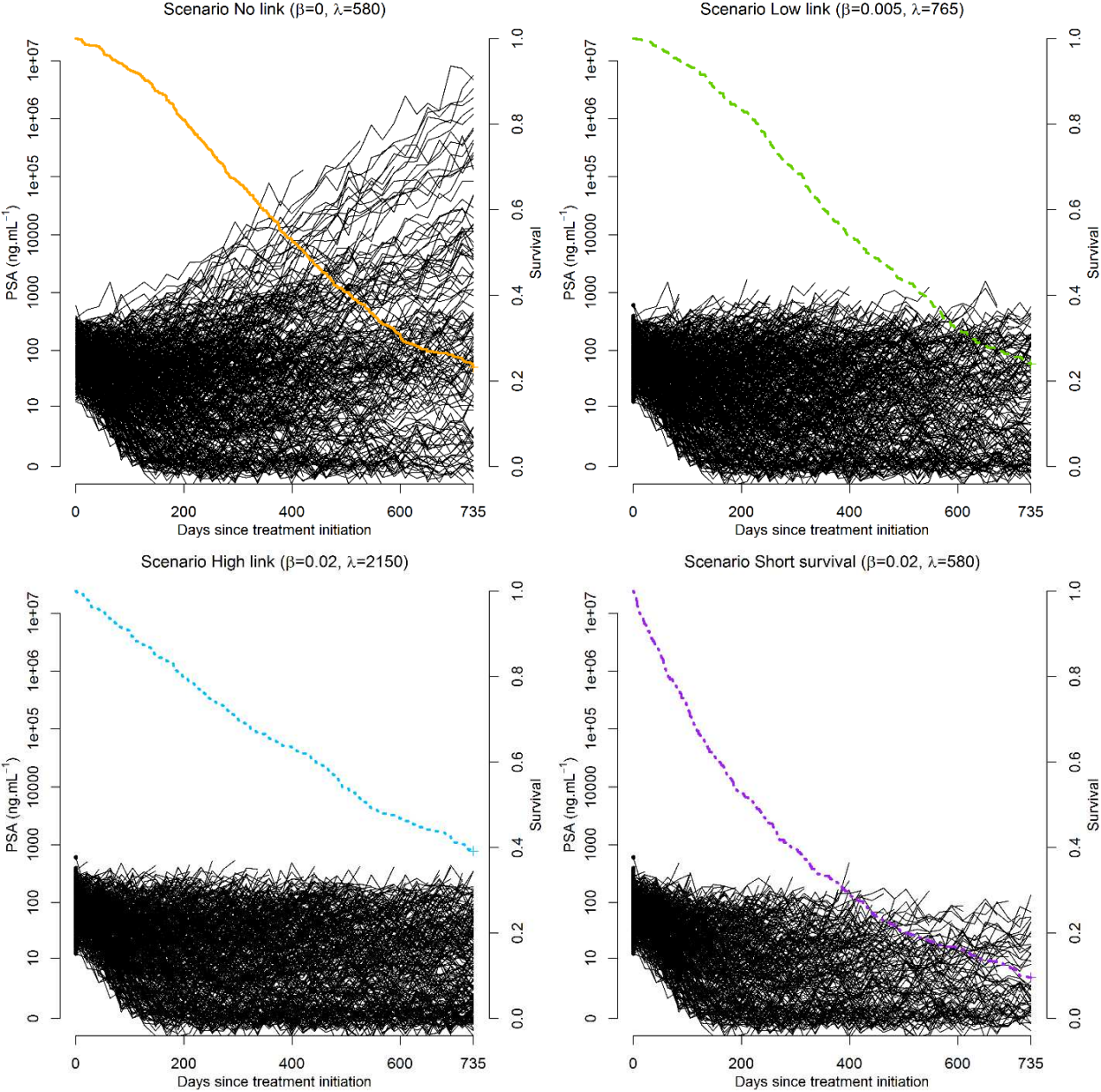

Figure 4: Boxplots of the relative estimation errors for parameters related to PSA for two-stage model (blue), joint sequential model (purple) and joint model (red), for the 4 scenarios. Top left: scenario No link ($\beta=0$, $\lambda=580$), top right: scenario Low link ($\beta=0.005$, $\lambda=765$), bottom left: scenario High link ($\beta=0.02$, $\lambda=2150$) and bottom right: scenario Short survival ($\beta=0.02$, $\lambda=580$).

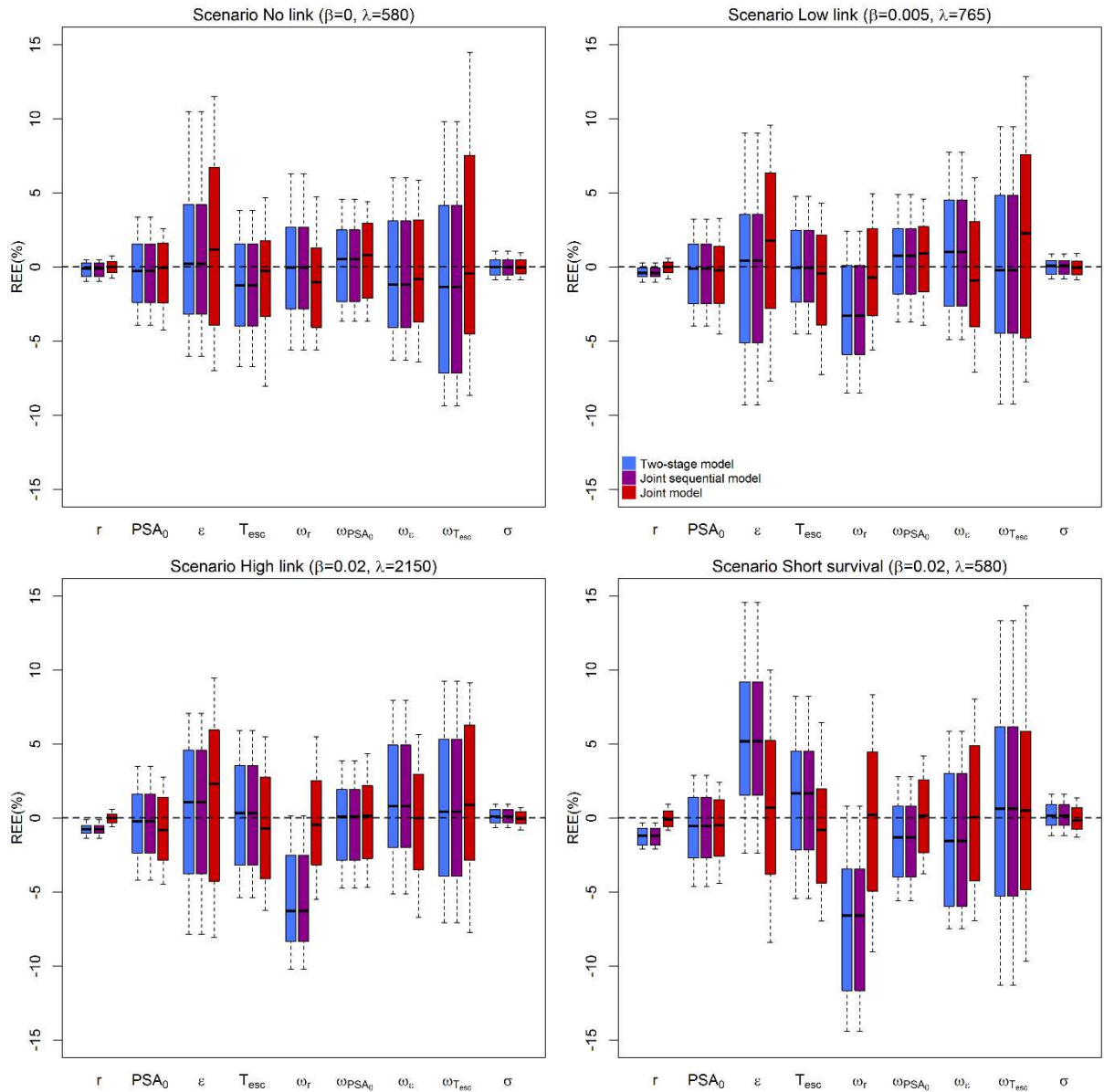

Figure 5: Boxplots of the relative estimation errors for parameters related to survival (in % except for β of the scenario No link for which estimated values are represented) for two-stage model (blue), joint sequential model (purple) and joint model (red), for the 4 scenarios. Top left: scenario No link (β=0, λ=580), top right: scenario Low link (β=0.005, λ=765), bottom left: scenario High link (β=0.02, λ=2150) and bottom right: scenario Short survival (β=0.02, λ=580).

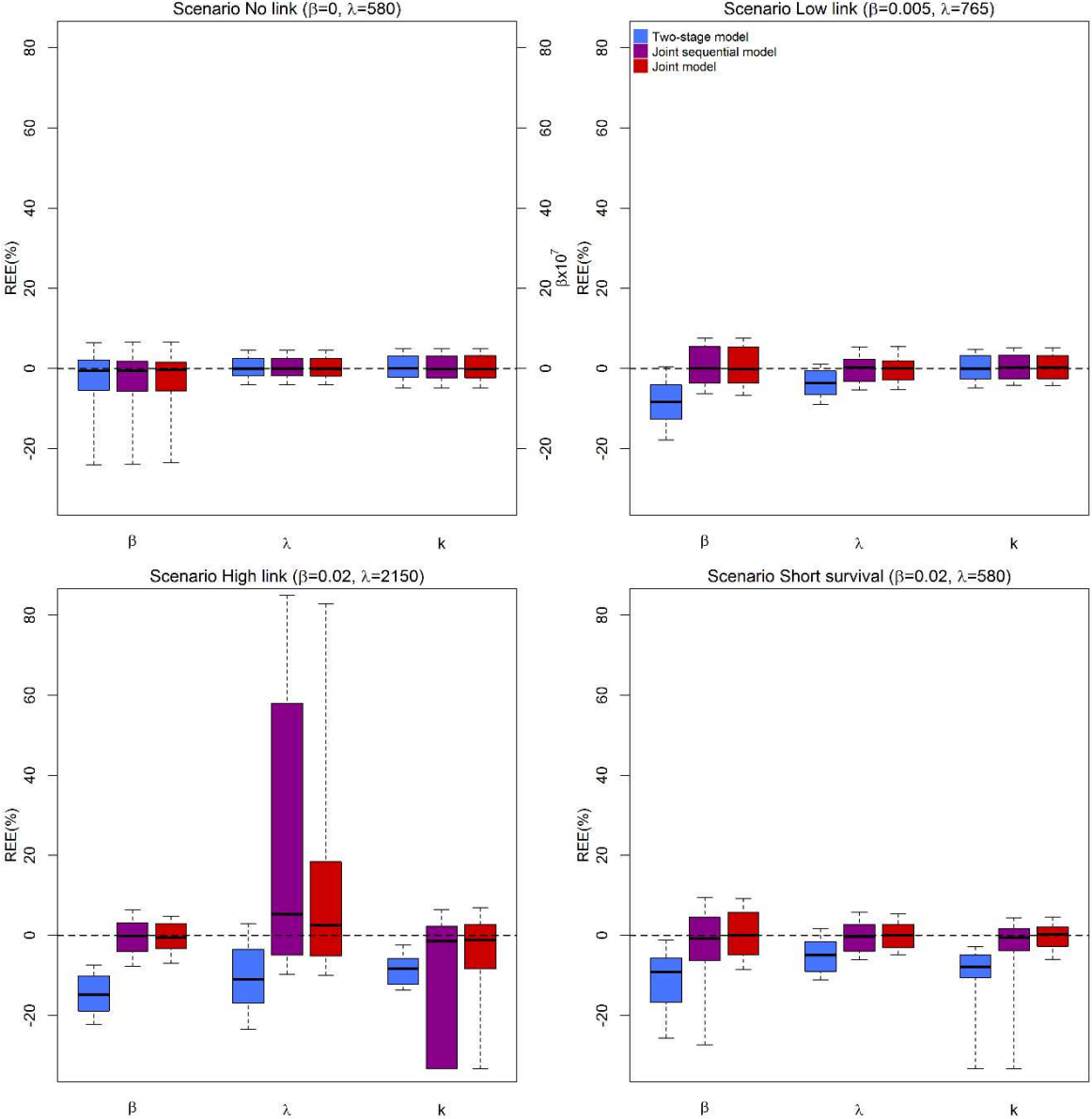

# Supplementary file 1: Impact of withdrawal from PSA follow-up on joint estimation

It is frequent in clinical trial protocols that treatment is stopped when a disease progression (increase in PSA or tumor size, adverse events) is observed. When treatment is stopped, the patient may drop out of the study. Although the vital status may continue to be collected, the PSA measurements are no longer recorded in the study.

### Simulation framework

For the sake of simplicity we consider here that a disease progression was only due to a PSA progression defined as *an increase of 25% above the nadir and of at least 2 ng/ml compared to nadir*. PSA data were removed accordingly in all datasets of the scenarios 'No link', 'High link' and 'Short survival' (see main analysis), leading to a dramatic decrease in the number of PSA measurements (Table S1).
In order to assess the impact of withdrawal data, two cases were considered:
- The vital status is known at the end of the study (scenarios 'Withdrawal')
- The vital status is censored at the time of the disease progression (scenarios 'Withdrawal + censor')

Thus, in the first case the number of observed deaths is equal to that observed in the main analysis, and in the second case the number of observed deaths is much smaller that observed in the main analysis (scenarios 'No withdrawal'). This results in much larger confidence interval for the Kaplan-Meier curves (Figure S1).

### Results

The variability of the parameter estimates increases when patients withdrawal from PSA follow-up (Figures S2 and S3) which was expected because there is a smaller number of PSA measurements. PSA kinetic parameters (Figure S2) were not affected by withdrawal, with or without censor of vital status. However all three parameters related to survival were estimated with a bias in case of censor of survival (Figure S3): $\beta$ and k tend to be overestimated while $\lambda$ tends to be underestimated. These trends led to an overestimation of the hazard function and hence an underestimation of the survival function.

## Conclusion

Drop out caused by predefined levels of PSA progression does not affect the estimation of the parameters associated with PSA kinetics.

Regarding survival parameters, a bias towards an overestimation of the hazard function may occur, in particular when not only PSA but also the vital status is not collected after disease progression.

Table S1: Number of PSA measurements per dataset and per patient and mean number of events per dataset in the total number of simulated datasets

|  |  | Scenario No link | | | Scenario High link | | | Scenario Short survival | | |
|---|---|---|---|---|---|---|---|---|---|---|
|  |  | No withdrawal | Withdrawal | Withdrawal + censor | No withdrawal | Withdrawal | Withdrawal + censor | No withdrawal | Withdrawal | Withdrawal + censor |
| Number of PSA measurements per dataset | | 11 104 | 3 062 | 3 062 | 12102 | 3488 | 3488 | 6 834 | 2 804 | 2 804 |
| Number of PSA measurements per patient | 1-5 | 7% | 65% | 65% | 8% | 64% | 64% | 29% | 70% | 70% |
| | 6-10 | 12% | 22% | 22% | 10% | 21% | 21% | 20% | 19% | 19% |
| | 11-20 | 26% | 7% | 7% | 22% | 7% | 7% | 26% | 7% | 7% |
| | 21-35 | 30% | 3% | 3% | 21% | 3% | 3% | 17% | 3% | 3% |
| | 36 | 24% | 1% | 1% | 39% | 4% | 4% | 8% | 1% | 1% |
| Mean number of events per dataset | | 379 | 379 | 66 | 305 | 305 | 36 | 462 | 462 | 145 |

Figure S1: Spaghetti-plots (black lines) and estimated Kaplan-Meier curves (colored solid lines) with their 95% confidence interval (colored dashed lines) for one typical dataset (N=500) for each scenario, without withdrawal (red curves), with withdrawal after PSA progression (pink curves) and with withdrawal and censor after PSA progression (grey curves).

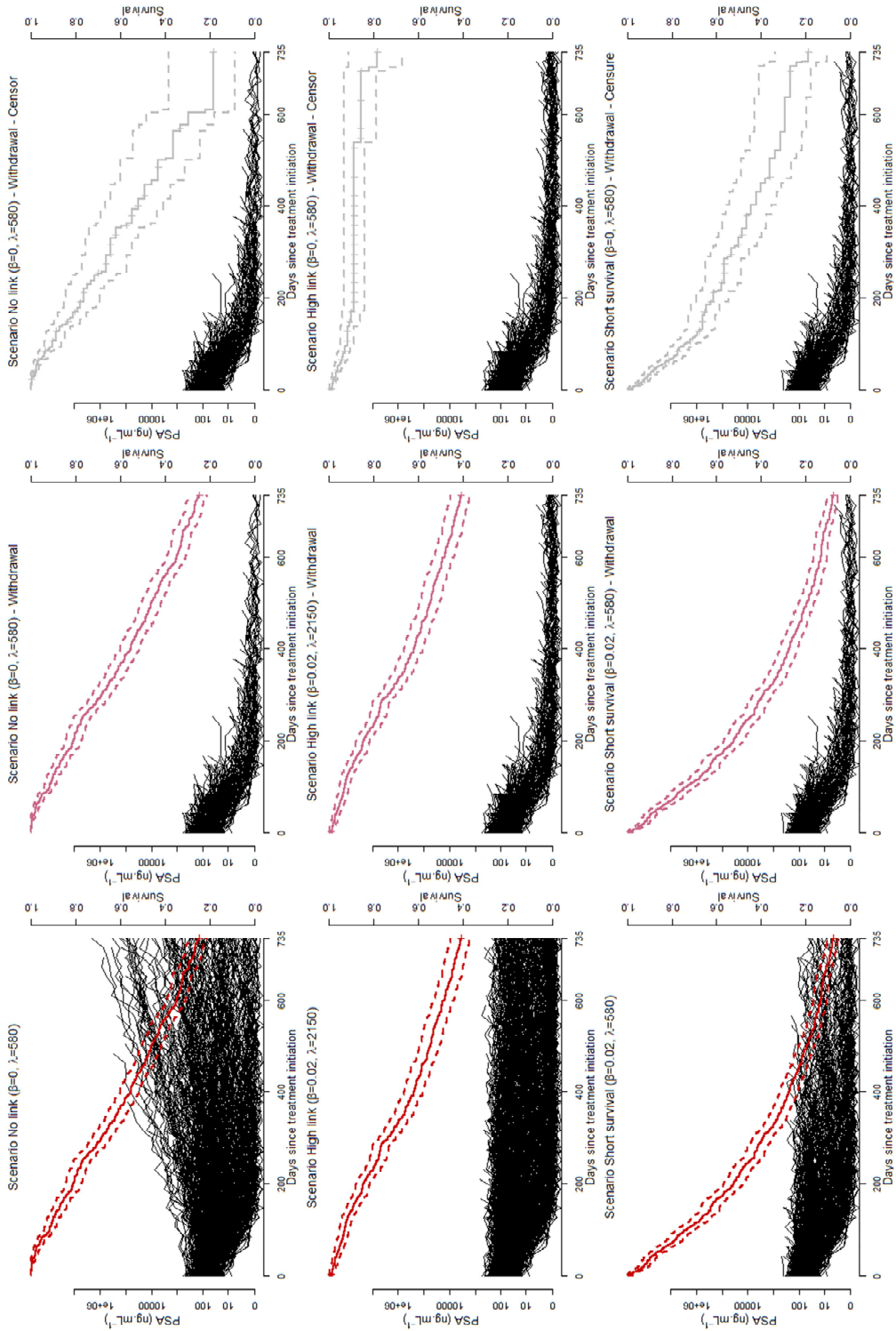

Figure S2: Boxplots of the relative estimation errors for the parameters related to PSA for joint model without withdrawal (red), with withdrawal (pink) and with withdrawal and censor (gray) and for the 3 scenarios. Top: scenario No link ($\beta=0$, $\lambda=580$), middle: scenario High link ($\beta=0.02$, $\lambda=2150$) and bottom: scenario Short survival ($\beta=0.02$, $\lambda=580$).

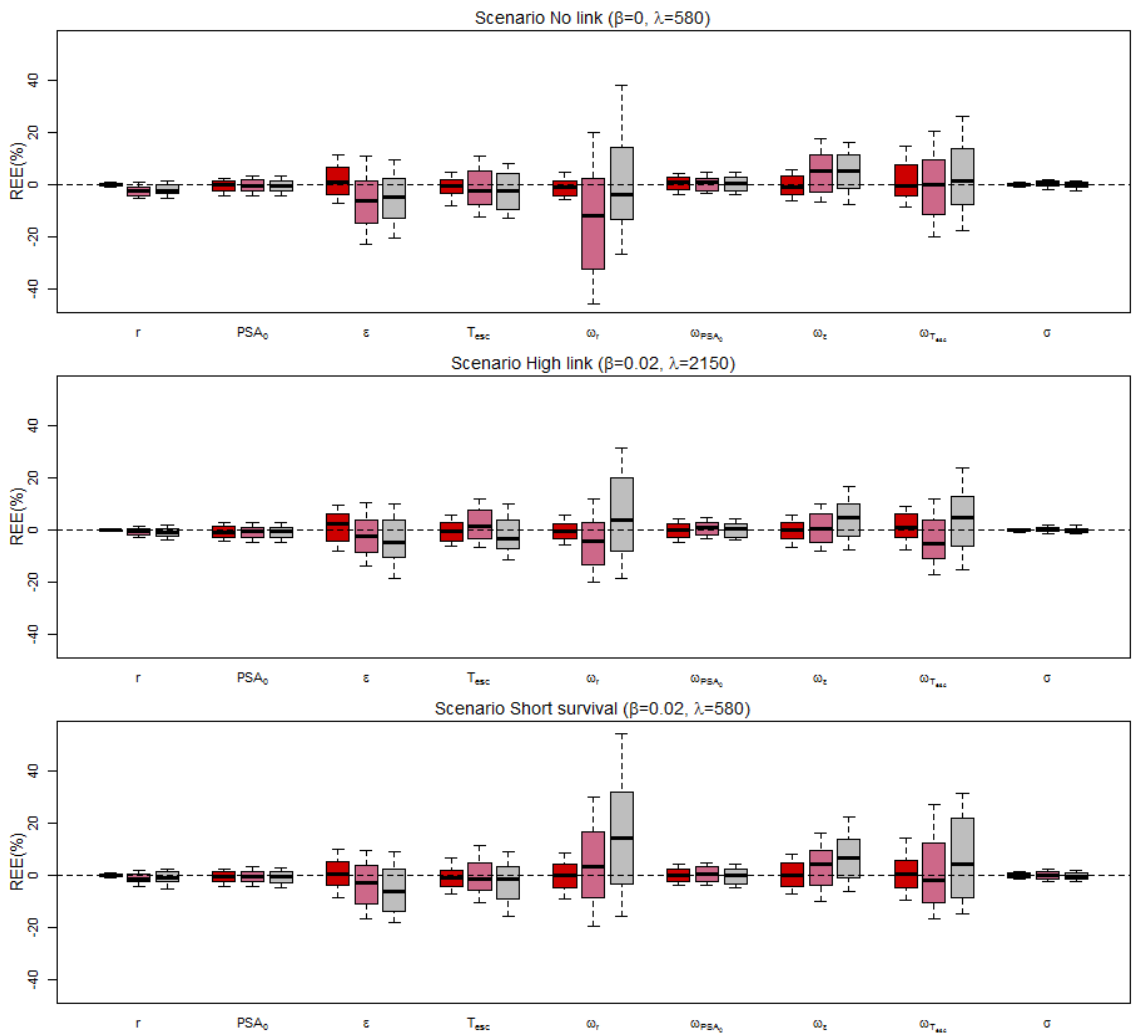

Figure S3: Boxplots of the relative estimation errors for the parameters related to survival (in % except for β of the scenario No link for which estimated values are represented) for joint model without withdrawal (red), with withdrawal (pink) and with withdrawal and censor (gray) and for the 3 scenarios. Top: scenario No link (β=0, λ=580), middle: scenario High link (β=0.02, λ=2150) and bottom: scenario Short survival (β=0.02, λ=580).

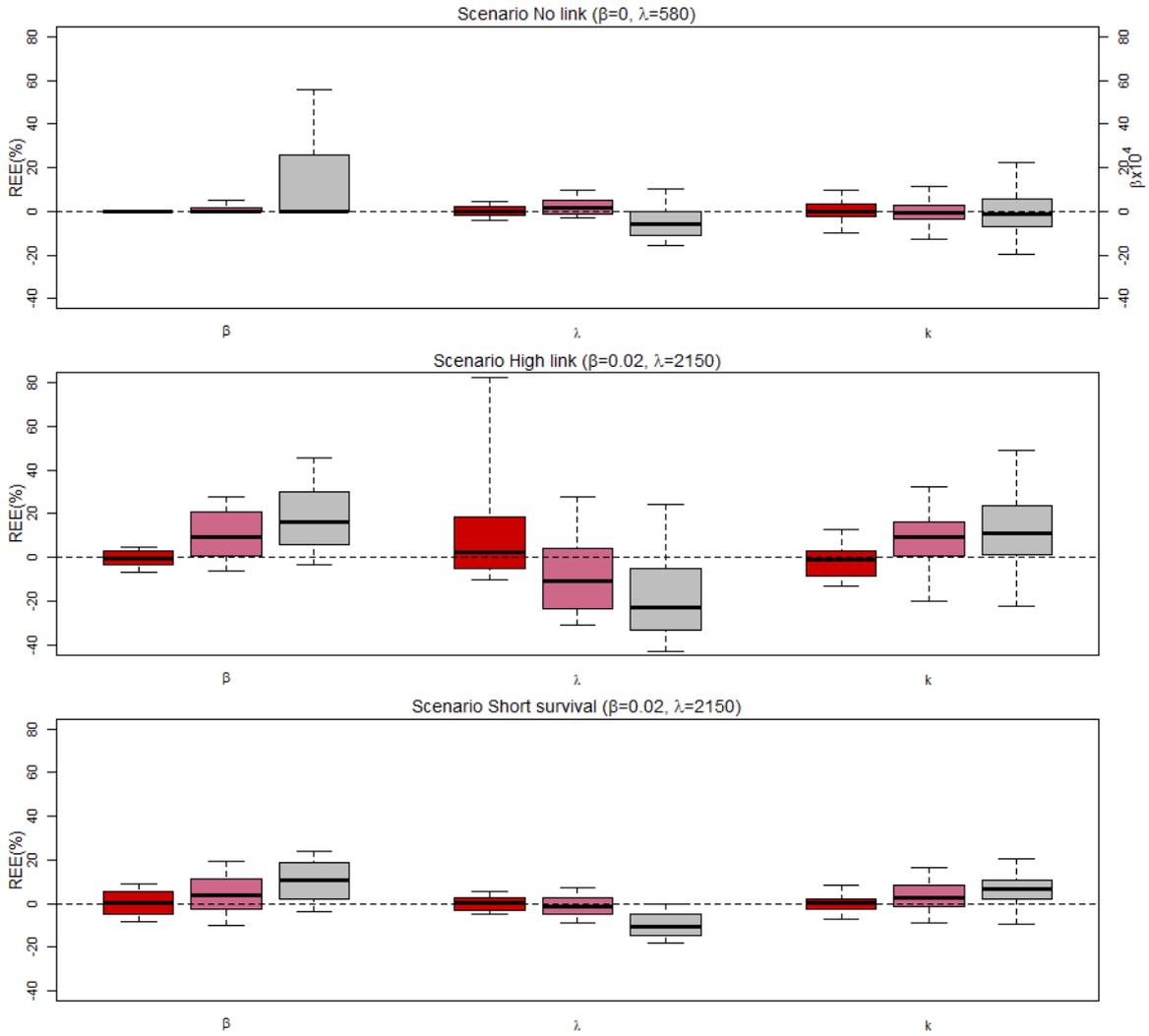